\documentstyle[12pt]{article}

\catcode`\@=11
\def\numberbysection{\@addtoreset{equation}{section}
        \def\theequation{\thesection.\arabic{equation}}}
\numberbysection
\def\beq{\begin{equation}}
\def\eeq{\end{equation}}
\def\IR{\relax{\rm I\kern-.18em R}}
\font\cmss=cmss10 \font\cmsss=cmss10 at 7pt
\def\IZ{\relax\ifmmode\mathchoice
{\hbox{\cmss Z\kern-.4em Z}}{\hbox{\cmss Z\kern-.4em Z}}
{\lower.9pt\hbox{\cmsss Z\kern-.4em Z}}
{\lower1.2pt\hbox{\cmsss Z\kern-.4em Z}}\else{\cmss Z\kern-.4em Z}\fi}
\def\inbar{\,\vrule height1.5ex width.4pt depth0pt}
\def\IB{\relax{\rm I\kern-.18em B}}
\def\IC{\relax\hbox{$\inbar\kern-.3em{\rm C}$}}
\def\ID{\relax{\rm I\kern-.18em D}}
\def\IE{\relax{\rm I\kern-.18em E}}
\def\IF{\relax{\rm I\kern-.18em F}}
\def\IG{\relax\hbox{$\inbar\kern-.3em{\rm G}$}}
\def\IH{\relax{\rm I\kern-.18em H}}
\def\II{\relax{\rm I\kern-.18em I}}
\def\IK{\relax{\rm I\kern-.18em K}}
\def\IL{\relax{\rm I\kern-.18em L}}
\def\IM{\relax{\rm I\kern-.18em M}}
\def\IN{\relax{\rm I\kern-.18em N}}
\def\IO{\relax\hbox{$\inbar\kern-.3em{\rm O}$}}
\def\IP{\relax{\rm I\kern-.18em P}}
\def\IQ{\relax\hbox{$\inbar\kern-.3em{\rm Q}$}}
\def\IGa{\relax\hbox{${\rm I}\kern-.18em\Gamma$}}
\def\IPi{\relax\hbox{${\rm I}\kern-.18em\Pi$}}
\def\ITh{\relax\hbox{$\inbar\kern-.3em\Theta$}}
\def\IOm{\relax\hbox{$\inbar\kern-3.00pt\Omega$}}

\def\Z{\IZ}
\def\Q{\IQ}
\def\R{\IR}
\def\C{\IC}

\def\om{\omega}
\def\del{\Delta}
\def\phim{\phi^{-1}}
\def\eps{\varepsilon}
\def\te{\otimes}
\def\id{{\rm id}}
\def\rc{\check{R}}
\def\dimq{{\rm dim}_{q}}
\def\lk{{\rm\ell k}}
\def\cf{{\cal F}}
\def\ca{{\cal A}}
\def\grc{{\rm Grc}(A)}
\def\hs{{\cal H}}
\def\dog{D^{\om}(G)}
\def\bea{\begin{eqnarray}}
\def\eea{\end{eqnarray}}
\def\qtr{{\rm tr}_{q}}
\def\tr{{\rm tr}}
\def\a{\alpha}
\def\b{\beta}
\def\vm{v^{-1}}
\def\T{{\cal T}}
\newcommand{\th}[1]{\theta_{#1}}
\newcommand{\ga}[1]{\gamma_{#1}}
\newcommand{\la}[1]{\omega_{#1}}

\newcommand{\elt}[2]
 { {\scriptstyle #1} | \! \raisebox{-.2ex}{\underline{\makebox[0.6em]{}}}
     \hspace{-0.6em}
     \raisebox{-2ex}{$\,\scriptstyle #2 \,$} }
\newcommand{\ele}[1]{\elt{#1}{e}}

\newcommand{\ord}{|G|}
\newcommand{\ordm}[1]{\ord^{-#1}}
\newcommand{\bo}[1]{\fbox{$#1$}}
\begin{document}
\begin{titlepage}
\begin{center}
{}~\hfill CERN-TH 6360/92\\
{}~\hfill CPT-91/PE.2634\\
{}~\hfill ENSLAPP-A-360/91\\
\vskip 0.6in
{\large \bf Quasi-Quantum Groups, Knots, Three-Manifolds, and Topological
Field Theory}
\vskip 0.5in
Daniel Altschuler\\
{\em CERN, Theory Division\\
     CH-1211 Gen\`{e}ve 23, Switzerland}
\vskip 0.2in
and
\vskip 0.2in
Antoine Coste
\footnote
{On leave from Centre de Physique Th\'{e}orique, CNRS Marseille-Luminy.} \\
{\em LAPP, Chemin de Bellevue, BP 110\\
     F-74941 Annecy-le-Vieux Cedex, France}
\end{center}
\vskip .5in
\begin{abstract}
We show how to construct, starting from a quasi-Hopf algebra, or
quasi-quantum group, invariants of knots and links.
In some cases, these invariants
give rise to invariants of the three-manifolds
obtained by surgery along these links.
This happens for a finite-dimensional quasi-quantum group,
whose definition involves
a finite group $G$, and a 3-cocycle $\om$,
which was first studied
by Dijkgraaf, Pasquier and Roche.
We treat this example in more detail, and argue
that in this case the invariants agree with
the partition function of the topological field theory of Dijkgraaf and Witten
depending on the same data $G, \,\om$.
\end{abstract}
\vskip .5 in
CERN-TH 6360/92\\
Revised version February 1992
\end{titlepage}
\roman{enumi}
\pagenumbering{arabic}
%
%
\section{Introduction}
It is by now well established that there are deep connections between
two-dimensional rational conformal field theories (RCFT), three-dimensional
topological field theories (TFT), and quantum groups
when $q$ is a root of unity, see e.g.
\cite{alvarezgaume,mooreseiberg,witten,ffk,aleks,gawedzki,rt1}.
Some aspects of this connection between RCFT, TFT, and quantum groups which
will
be important in the sequel are: i) the imaginary exponentials of
the conformal weights of primary fields in RCFT, are equal to the values
on certain representations
of a central element $v$ in the quantum group, playing the role of a Casimir
operator, ii) the elements of the matrix $S$ of modular transformations
on the torus are given by the trace of an expression involving the
$R$\/-matrix acting in tensor products of representations
- in particular Verlinde's quantum dimensions agree with quantum dimensions
as defined for quantum groups, and fusion rules are given by the
``truncated'' tensor products of representations of quantum groups
\cite{lotsofguys},
iii) the representation of the braid group arising as the monodromy of
the chiral blocks \cite{ktk}
is equivalent to the representation of the braid group
coming from $R$\/-matrices, etc\ldots

A key element in any attempt at understanding these coincidences is
the fact that both RCFT and quantum groups are sources of
topological invariants of knots, links and three-dimensional manifolds
(through the TFT reinterpretation of RFCT). For instance, the invariants
of the Hopf link are the elements of the matrix $S$
\cite{alvarezgaume,mooreseiberg},
and consideration of a chain of three circles is the key to proving
Verlinde's formula.
The construction of invariants of links from the representation theory
of quantum groups was developed in \cite{kr,reshet,rt2}. In its most
general form it appears in \cite{rt2}, where the concept of ribbon Hopf
algebras is introduced. Examples of ribbon Hopf algebras are the
``usual'' quantum groups \cite{fathers}
$U_{q} {\cal G}$ where ${\cal G}$ is a
semi-simple Lie algebra \cite{rt1}, the double $D(G)$
of a  finite group $G$, and many more are discussed
in a recent paper of Kauffmann and Radford \cite{kauffrad}.
To our taste, the above coincidences
are best explained in \cite{rt2}, where
a TFT, formalized in the sense of Atiyah and Segal \cite{atiyah},
is reconstructed from ribbon Hopf algebras of a particular class
called modular Hopf algebras by these authors. Roughly speaking, a
modular Hopf algebra $A$ is a ribbon Hopf algebra with a finite set
of representations which is closed under the tensor product operation,
up to representations of quantum dimension zero;
$U_{q} {\cal G}$ for $q$ a root of unity \cite{rt1,kirmel,tv}
and $D(G)$ \cite{alcos} belong to this class.

In another direction, one may ask how to construct canonically a
quantum group, starting from a TFT. Already in the work
of Moore and Seiberg \cite{ms}, it is clear that
this problem is analogous to
the Tannaka-Krein reconstruction of a group $G$ from
a category of vector spaces
which at the end, become representations of $G$.
In his work, Majid \cite{majid}
solves the problem, showing that the initial data
is the category of cobordisms
instead of a category of vector spaces. We find it worthwhile to
explain briefly what is this category, in order to be able
to summarize Majid's result.
The category of cobordisms has as objects
two-dimensional oriented Riemann surfaces $\Sigma$,
and morphisms are three-dimensional compact
manifolds $M$ with given boundary Riemann surfaces
$\partial M = \Sigma^{+} \cup \Sigma^{-}$,
a morphism being from
the components $\Sigma^{+}$ of the boundary
with positive orientation to those
with a negative orientation, $\Sigma^{-}$.
The Atiyah-Segal modular functor which
defines a TFT is $\Sigma \mapsto \hs_{\Sigma}$,
$M \mapsto {\cal O}_{M}$, where
$\hs_{\Sigma}$ is the Hilbert space of TFT
associated to $\Sigma$, which is
nothing but the space of conformal blocks on
$\Sigma$, and ${\cal O}_{M}$ is a
linear map $\hs_{\Sigma^{+}} \rightarrow \hs_{\Sigma^{-}}$.
Then $\langle \psi^{\prime} | {\cal O}_{M} | \psi \rangle$
is the amplitude for ``propagation'' from the initial state
$\psi\in\hs_{\Sigma^{+}}$ to the final state
$\psi^{\prime} \in\hs_{\Sigma^{-}}$.
Now try to define the quantum group $A$
associated with this TFT to be the vector space of functions
$a : \Sigma \mapsto a_{\Sigma} \in {\rm End} \hs_{\Sigma}$,
such that
${\cal O}_{M} \, a_{\Sigma^{+}} = a_{\Sigma^{-}} \, {\cal O}_{M}$.
This space becomes an algebra with the
obvious product, and one can easily define
also a coproduct by considering functions on $\Sigma \cup \Sigma^{\prime}$
(disjoint union). The trouble is that, as pointed out in \cite{majid}, in
general this coproduct $\del$ will fail to be coassociative, it will be
quasi-coassociative:
\beq
(\id \te \del) (\del(a)) = \phi \, (\del \te \id) (\del(a)) \phim,
\label{quasi}
\eeq
where $\phi \in A \te A \te A$, and satisfies natural pentagon and hexagon
identities (there is also a natural $R$\/-matrix). This kind of object,
now called quasi-Hopf algebra, was
invented by Drinfeld \cite{drinfeld} some time before, but with a
completely different motivation, which we explain below. We should mention
at this point that the relevance of quasi-Hopf algebras for TFT could have
been foreseen in the paper of Dijkgraaf, Pasquier and Roche \cite{dpr,roche},
where they built an interesting example $\dog$, which is a ``deformation'' of
$D(G)$ involving a non-trivial 3-cocycle $\om$ of $G$,
in order to reproduce the
fusion rules of the Dijkgraaf-Witten TFT \cite{dvvv,dw}
defined with the same data $G, \om$. Mack and Schomerus \cite{mascho}
have proposed to use quasi-Hopf algebras in RCFT, e.g. to reproduce
the primary field content and fusion rules of the Ising model.
To achieve this,
however, they seem to need to generalise even more the quantum groups, as
witnessed by their definition of weak quasi-Hopf algebras.

Drinfeld's motivation, as far as we know,
was based on the observation that when
one tries to deform the coproduct $\del$ of a Hopf algebra, setting:
\[
\del^{f}(a) = f \, \del(a) \, f^{-1},
\]
with $f \in A \te A$ an invertible element,
then $\del^{f}$ is no longer coassociative,
but satisfies (\ref{quasi}) above, where
\[
\phi = f_{23} \, (\id \te \del)(f) (\del \te \id)(f^{-1}) \, f^{-1}_{12}.
\]
Here and later, $f_{ij}$ means $f$ acting
non-trivially in the $i$\/-th and $j$\/-th
place of $A \te A \te A$. Now if one defines quasi-Hopf algebras by the
property (\ref{quasi}), one gets a class of objects which is stable
under the mapping $\del \rightarrow \del^{f}$, called ``twist by f''.
This twist takes $\phi$ into
\[
\phi^{f} = f_{23} \, (\id \te \del)(f) \, \phi \,
  (\del \te \id)(f^{-1}) \, f^{-1}_{12}.
\]
Twists also preserve the class of quasitriangular quasi-Hopf algebras,
which will be defined in sect. 2. Drinfeld proved that all quasitriangular
quasi-Hopf algebras, which are quantum deformations of universal enveloping
algebras, can be obtained by twisting $U_{q}{\cal G}$. (Note that this
result does not apply to $\dog$.) He gave a very interesting example
$A_{{\cal G}, t}$ of
such a quasi-Hopf deformation of $U {\cal G}$, the universal enveloping
algebra. The algebra and coalgebra structures of
$A_{{\cal G}, t}$ are the same
as those of $U {\cal G}$, but he imposes quasitriangularity in the
quasi-Hopf sense, where $R = \exp ht$, $h$ is a deformation parameter, and
$t \in {\cal G} \te {\cal G}$ is a ${\cal G}$\/-invariant symmetric tensor,
e.g. the tensor coming from the Cartan-Killing form. The element $\phi$ is
found to be completely determined by the monodromy of the 4-point functions
solving the Knizhnik-Zamolodchikov equations. Since the space of solutions
of these equations affords a representation of the braid group, this gives
a natural explanation for the coincidence of braid group representations
coming from RCFT and quantum groups, which was mentioned before.

In this paper we present a natural
extension of the constructions of Reshetikhin and Turaev \cite{rt1,rt2}
to the case of quasi-Hopf algebras.
More precisely, for any ribbon quasi-Hopf algebra we define regular
isotopy invariants of coloured ribbon graphs, the colours being
finite-dimensional representations. This result is very general and
can be applied to a much broader set of algebras and topological setups
than those considered later in the paper, for instance to the construction
of ambient isotopy invariants of links. We intend to explore some of these
questions in a future work. Our motivation for constructing these
ribbon invariants was to be able to understand the
topological field theory of Dijkgraaf and Witten \cite{dw},
whose theory was further investigated by Freed and Quinn \cite{freed},
in the case of a non-trivial cocycle
$\om$, using only the algebra $\dog$ of \cite{dpr}. We succeeded in
finding a 3-manifold invariant, considering surgery on the ribbon graphs
coloured by a representation of $\dog$, which in the examples that we
computed explictly, coincides with the invariant of \cite{dw}. We
conjecture that this holds in general. One advantage of our
approach for constructing the invariants is that it lends
itself to practical computation
from a surgery presentation of the manifold, whereas
the original definition requires the knowledge of a triangulation, which
is generally more difficult to find.

In section 2, we
recall the basic definitions from Drinfeld's \cite{drinfeld} original
papers. In section 3, we give an important theorem on the square of the
antipode in quasi-Hopf algebras possessing an $R$\/-matrix, generalizing
a theorem of Drinfeld \cite{dr2} for Hopf algebras.
In section 4 we define invariants of ribbon graphs, which are framed
links (tangles) with some open ends. These invariants are intertwining
operators for a ribbon quasi-Hopf algebra. In the particular case of graphs
with only closed ribbons (annuli), these invariants are pure numbers and
similar to the Reshetikhin-Turaev version of Jones's polynomial.
In section 5 we first prove that the algebra $\dog$ is a ribbon quasi-Hopf
algebra, and then we show that it even allows to define
invariants of 3-manifolds using surgery. In some simple cases we compute
these invariants, checking the properties predicted by our conjecture.
\section{Definitions}
Let $A$ be an associative algebra over $\C$,
with a unit element $1$. We say that $A$ is a quasi-bialgebra if there are
algebra homomorphisms
$\del : A \rightarrow A \te A$, $\eps : A \rightarrow \C$
and an invertible $\phi\in A \te A \te A$, such that:
\beq
(\id \te \del) (\del(a)) = \phi \, (\del \te \id) (\del(a)) \phim
\;\;\; a \in A
\label{quasiass}
\eeq
\beq
(\id \te \id \te \del)(\phi) \, (\del \te \id \te \id)(\phi) =
(1 \te \phi) \, (\id \te \del \te \id)(\phi) \, (\phi \te 1)
\label{pentagon}
\eeq
\beq
(\eps \te \id) \circ \del = \id = (\id \te \eps) \circ \del
\label{counit}
\eeq
\beq
(\id \te \eps \te \id) (\phi) = 1
\label{triangle}
\eeq
The map $\del$ is called the coproduct, and $\eps$ the counit.

Let us briefly recall some of the main
consequences of these definitions in the representation theory of $A$.
In this paper we will be dealing only with finite-dimensional
representations $(\pi, V)$ of $A$, which consist of
a finite-dimensional vector space $V$ over $\C$, and
a representation $\pi : A \rightarrow {\rm End} V$.
We will also use the equivalent definition of
an $A$-module V, and write $a \cdot v$ for $\pi(a) v$, $a\in A$, $v\in V$.
Given two such representations $(\pi_{1}, V_{1})$ and
$(\pi_{2}, V_{2})$ one may construct
representations $(\pi_{12}, V_{1}\te V_{2})$ and $(\pi_{21}, V_{2}\te V_{1})$
by setting
\beq
\pi_{12} = (\pi_{1}\te \pi_{2}) \Delta
\label{v1v2}
\eeq
and similarly for $\pi_{21}$. Suppose we are given three representations
$(\pi_{i}, V_{i})$, $i=1,2,3$. Set
\beq
\phi^{V_{1}, V_{2}, V_{3}} = (\pi_{1} \te \pi_{2} \te \pi_{3})(\phi).
\eeq
Then (\ref{quasiass}) says that $\phi^{V_{1}, V_{2}, V_{3}} :
(V_{1} \te V_{2}) \te V_{3} \rightarrow V_{1} \te (V_{2} \te V_{3})$
is an intertwiner, and therefore the representations (modules)
$(V_{1} \te V_{2}) \te V_{3}$ and $V_{1} \te (V_{2} \te V_{3})$ are equivalent.
Now take four representations. The identity (\ref{pentagon}) implies that
the diagram
\[
\begin{array}{ccccc}
((V_{1} \te V_{2}) \te V_{3}) \te V_{4} & \rightarrow &
(V_{1} \te V_{2}) \te (V_{3} \te V_{4}) & \rightarrow &
V_{1} \te (V_{2} \te (V_{3} \te V_{4})) \\
\downarrow & & & & \downarrow\\
(V_{1} \te (V_{2} \te V_{3})) \te V_{4} & & \longrightarrow & &
V_{1} \te ((V_{2} \te V_{3}) \te V_{4})
\end{array}
\]
commutes, where the arrows are $\phi^{V_{1} \te V_{2}, V_{3}, V_{4}}$,
$\phi^{V_{1}, V_{2}, V_{3} \te V_{4}}$, etc. This explains the use
of the name {\em pentagon identity} for eq. (\ref{pentagon}).

Using the counit $\eps$, one obtains a one-dimensional representation of $A$ on
$\C$. Then (\ref{counit}) means that
\[ V \te \C = V = \C \te V \]
for any $A$\/-module $V$. We will refer to $(\eps, \C)$ as the trivial
representation.
One sees that (\ref{pentagon}) and (\ref{triangle})
together imply
\beq
(\eps \te \id \te \id) (\phi) = (\id \te \id \te \eps)(\phi) = 1,
\label{tri}
\eeq
therefore in a tensor product of three representations one may forget a
trivial factor.

A quasi-bialgebra $A$ is called a quasi-Hopf algebra if there exists an
antiautomorphism $S$ of $A$, i.e. $S(ab) = S(b) S(a)$, and two elements
$\alpha, \beta \in A$ such that:
\beq
\sum_{i} S(a_{i}^{(1)}) \alpha a_{i}^{(2)} = \eps(a) \alpha
\label{anti1}
\eeq
\beq
\sum_{i} a_{i}^{(1)} \beta S(a_{i}^{(2)}) = \eps(a) \beta
\label{anti2}
\eeq
for $a \in A$ and $\sum_{i} a_{i}^{(1)} \te a_{i}^{(2)} = \del(a)$, and
\beq
\sum_{i} X_{i} \beta S(Y_{i}) \alpha Z_{i} = 1, \;\;\;{\rm where} \;
\sum_{i} X_{i} \te Y_{i} \te Z_{i} = \phi,
\label{snake1}
\eeq
\beq
\sum_{j} S(P_{j}) \alpha Q_{j} \beta S(R_{j}) = 1, \;\;\;{\rm where} \;
\sum_{j} P_{j} \te Q_{j} \te R_{j} = \phim.
\label{snake2}
\eeq
We note the following two consequences of the definitions of $S, \alpha,
\beta$:
\beq
\eps(\alpha)\eps(\beta) = 1,
\label{epsalphabeta}
\eeq
\beq
   \eps \circ S = \eps.
\eeq
The map $S$ is called the antipode.
It allows us to define the dual representation
$(\pi^{*},V^{*})$ of $(\pi, V)$, where $V^{*}$ is the dual space, by
\beq
\pi^{*}(a) = (\pi \circ S(a))^{t},
\eeq
the superscript $t$ denoting the transposed map.

In the theory of Hopf algebras,
the following relation is well-known:
\[
\del(a) = (S \te S) (\del^{\prime} \circ S^{-1}(a)),
\]
where $\del^{\prime}= \sigma \circ \del$, $\sigma : a \te b \mapsto b \te a$.
Later on we will need the generalization of this, which is due to Drinfeld. Let
\beq
\sum_{j} A_{j} \te B_{j} \te C_{j} \te D_{j}
  = (\phi \te 1) (\del \te \id \te \id)(\phim),
\eeq
\beq
\gamma = \sum_{j} S(B_{j}) \alpha C_{j} \te S(A_{j}) \alpha D_{j},
\eeq
\beq
\sum_{i} K_{i} \te L_{i} \te M_{i} \te N_{i}
  = (\del \te \id \te \id)(\phi) (\phim \te 1),
\eeq
\beq
\delta = \sum_{i} K_{i} \beta S(N_{i}) \te L_{i} \beta S(M_{i}).
\eeq
Then for any $a\in A$,
\beq
f \del(a) f^{-1} = (S \te S) (\del^{\prime} \circ S^{-1}(a))
\label{eqt}
\eeq
where
\beq
f = \sum_{i} (S \te S)(\del^{\prime}(P_{i}))
  \cdot \gamma \cdot \del(Q_{i} \beta S(R_{i})).
\label{eff}
\eeq
Moreover,
\beq
\gamma = f \, \del(\alpha), \;\;\; \delta = \del(\beta) \, f^{-1}.
\label{prof}
\eeq
In fact, Drinfeld shows that $f$ defines a twist of $A$, where the modified
coproduct is the r.h.s. of (\ref{eqt}).

A quasi-Hopf algebra is termed {\em quasitriangular},
if there exists an invertible
element \mbox{$R \in A \te A$}, such that:
\beq
\del^{\prime}(a) = R \, \del(a) \, R^{-1}
\label{deltaprime}
\eeq
\beq
(\del \te \id)(R) = \phi_{312} R_{13} \phim_{132} R_{23} \phi,
\label{quasitriangle1}
\eeq
\beq
(\id \te \del)(R) = \phim_{231} R_{13} \phi_{213} R_{12} \phim,
\label{quasitriangle2}
\eeq
where we have used the following notation:
$R_{ij}$ means $R$ acting non-trivially in the
$i$\/-th and $j$\/-th slot of $A \te A \te A$.
If $s$ denotes a permutation of
$\{1,2,3\}$ and $\phi = \sum_{i} a_{i}^{1} \te a_{i}^{2} \te a_{i}^{3}$
then we set
$\phi_{s(1)s(2)s(3)} =
\sum_{i} a_{i}^{s^{-1}(1)} \te a_{i}^{s^{-1}(2)} \te a_{i}^{s^{-1}(3)}$.
 From these relations one deduces the quasi-Yang-Baxter equation:
\beq
R_{12} \phi_{312} R_{13} \phim_{132} R_{23} \phi
 = \phi_{321} R_{23} \phim_{231} R_{13} \phi_{213} R_{12}.
\label{qyb}
\eeq
The translation of (\ref{quasitriangle1}) and (\ref{quasitriangle2}) in the
language of commutative diagrams leads to hexagons \cite{drinfeld}.
The following property of $R$ can be derived easily:
\beq
 ( \eps \te \id ) R = ( \id \te \eps ) R = 1.
\eeq
The most significant consequence of (\ref{deltaprime}) in representation theory
is that the representations $(\pi_{12}, V_{1} \te V_{2})$ and
$(\pi_{21}, V_{2} \te V_{1})$ are equivalent:
\beq
\pi_{21} (a) = \rc_{12} \, \pi_{12}(a) \, \rc^{-1}_{12}
\eeq
where $\rc_{12} : V_{1} \te V_{2} \rightarrow V_{2}\te V_{1}$ is given by
$\rc_{12} = P_{12} (\pi_{1} \te \pi_{2}) R$ and $P_{12}$ is the operator which
permutes the vectors in $V_{1}$ and $V_{2}$.
\section{The square of the antipode}
Let A be a quasi-Hopf algebra with an $R$\/-matrix satisfying
(\ref{deltaprime}).
Generalizing a theorem of Drinfeld for Hopf algebras, we will prove that
for any $a \in A$,
\beq
S^2(a) = uau^{-1},   \label{S2}
\eeq
where $u$ is given by the formula:
\beq
u = \sum_{j,p} S(Q_j \beta S(R_j)) \,  S(b_p) \alpha a_p P_j,
\eeq
in terms of
\beq
R = \sum_p a_p \te b_p , \ \
\phim = \sum_j P_j \te Q_j \te R_j .
\eeq
Let us start by showing that:
\beq
  S^2 (a)u = u a. \label{S2u}
\eeq
Set $ (\del  \te \id )\del (a) = \sum_k f_k \te g_k \te h_k $ ; using
(\ref{counit}) and (\ref{anti1}) one has
\beq
\sum_k S(f_k) \alpha g_k \te h_k = \alpha \te a,
\eeq
and therefore
\beq
S^2(a)u = \sum_{j,k,p} S^{2}(h_k) S(Q_j \beta S(R_j) ) \,
 S(b_p)S(f_k)\alpha     g_k a_p P_j .
\eeq
But (\ref{deltaprime}) implies
\beq
\sum_{k,p} a_p f_k \te b_p g_k \te h_k =
               \sum_{k,p} g_k a_p \te f_k b_p \te h_k,
\eeq
so that:
\beq
S^{2} (a) u =\sum_{j,k,p} S(g_k Q_j \beta S(h_k R_j) ) S(b_p)\alpha a_p
                                                        f_k P_j.
\eeq
Now $ (\del \te \id )\del (a)\phim =\phim (\id \te \del)\del (a)$,
(\ref{counit}) and (\ref{anti2}) lead to (\ref{S2u}).

\noindent
Our next move is to establish the lemma:
\beq
S(\alpha) u = \sum_{p} S(b_p) \alpha a_p. \label{lemmaSalphau}
\eeq
To prove it, one performs in $u$ the substitution
$$ \sum_j P_j \te Q_j \te R_j \te 1
 = (\del \te \id \te \id)(\phim) (\id \te \id \te \del )(\phim)
                  (1 \te \phi) (\id \te \del \te \id )(\phi) $$
and  simplifies in several steps the resulting expression for
$ S(\alpha) u $ by use of
(\ref{triangle}), (\ref{tri}), (\ref{anti1}) and (\ref{anti2}).

\noindent
Now (\ref{lemmaSalphau}) implies
\beq  ut = \alpha  \label{uteqalpha} \eeq
where we set:
\beq
  t =\sum_q S^{-1}(\alpha d_q )c_q,\;\;\; R^{-1} = \sum_q c_q \te d_q
\eeq
Plugging (\ref{uteqalpha}) into (\ref{snake2}) gives
\beq
 1 = \sum_{j} S(P_j) ut Q_j \beta S(R_j)
   = u \sum_{j} S^{-1}(P_j) t Q_j \beta S(R_j)
   = S^{2}(\sum_{j} S^{-1}(P_j) t Q_j \beta S(R_j)) \, u
\eeq
Therefore u, which has both a left and right inverse, is invertible, and
$ S(u) $ too. This completes the proof of (\ref{S2}).
Some straightforward corollaries are:
\begin{enumerate}
\item $S^2 (u)=u$
\item the element $uS(u)=S(u)u$ is central
\item
$\sum_{p} S(b_p)\alpha a_p = S(\alpha)u = S(t)S(u)u
    =S(u)u \ \sum_{q} S(c_q)\alpha d_q$.
\end{enumerate}
\noindent
Notice also that (\ref{triangle}) and (\ref{epsalphabeta}) ensure
$ \eps(u)=1 $.

The most important consequence of this theorem for representation theory,
is that for any quasitriangular quasi-Hopf algebra $A$, and for any
finite-dimensional representation $(\pi,V)$ of $A$, the double dual
$(\pi^{**},V^{**})$ is equivalent to $(\pi,V)$, the intertwiner being
$\pi(u)$. This means also that the (right) dual $(\pi^{*},V^{*})$ is equivalent
to the {\em left dual} representation $(\mbox{}^{*}\pi, V^{*})$ which is
defined \cite{drinfeld} by
$\mbox{}^{*}\pi(a) = (\pi \circ S^{-1}(a))^{t}$ for $a\in A$.
\section{The generalized Reshetikhin-Turaev functor}
\subsection{Ribbon quasi-Hopf algebras}
Let $A$ be a quasitriangular quasi-Hopf algebra. We propose the
following generalisation of the notion of ribbon Hopf algebra of Reshetikhin
and Turaev. We say that $A$ is a ribbon quasi-Hopf algebra, if there exists
a central element $v \in A$ such that
\begin{description}
\item [R1.] $v^{2} = u S(u)$
\item [R2.] $S(v) = v$
\item [R3.] $\eps(v) = 1$
\item [R4.] $\del(uv^{-1}) = f^{-1} ((S \te S)(f_{21})) (uv^{-1} \te uv^{-1}),$
\end{description}
where $f$ is defined in (\ref{eff}). We shall comment later on the
consequence of these conditions, and give a detailed example of
ribbon quasi-Hopf algebra.
\subsection{Coloured ribbon graphs}
A ribbon graph \cite{rt2} can be defined
as a regular projection on a plane of a
finite set of oriented ribbons in $\R^{3}$, i.e. two-dimensional oriented
manifolds with boundaries which are the
images of non self-intersecting smooth
embeddings $ [0,1] \times [0,1] \rightarrow \R^{3} $ (open ribbons)
or $ S^{1} \times [0,1] \rightarrow \R^{3} $ (annuli).
Note that Moebius strips are excluded by this definition so that
ribbons have a ``white'' and a ``black'' side.
The definition of ribbon graphs also
assumes that the white side is always facing the observer on the top and bottom
of the figure. Furthermore the extremities of all the open ribbons are
vertical. Ribbons are also directed, i.e. equipped with an arrow.
An example of ribbon graph is shown on figure 1.

Two graphs are considered equivalent if and only if they are projections of
isotopic ribbons. Here by isotopy we mean a smooth isotopy of $\R^{3}$ which
preserves the directions of arrows, the orientation of the graph surface,
and keeps the ends of open ribbons fixed.
For convenience we will represent pictorially such a ribbon graph as the
projection of an oriented link (with possibly open components). This means that
we identify the graphs as in figure 2.

Now we define {\em coloured ribbon graphs}, or c-graphs for short. Let $A$ be
a ribbon quasi-Hopf algebra.
Denote by $N(A)_{k}$ the class of all words (formal non-associative
expressions)
of the form
\beq
((((V_{1}^{\eps_{1}} \Box (( V_{2}^{\eps_{2}} \Box \cdots
)) \cdots ) \Box V_{k}^{\eps_{k}})) )
\eeq
where the $k$ letters $V_{i}$ are $A$\/-modules,
$\eps_{i} = \pm 1$, and $V^{1} = V$,
$V^{-1} = V^{*}$. There is no restriction on the
location of parentheses, but we
regard two words with the same letters but a different
distribution of parentheses
as being distinct,
e.g. $(V_{1} \Box V_{2}) \Box V_{3} \neq V_{1} \Box (V_{2} \Box V_{3})$.
By definition $N(A)_{0}$ consists of the single word
$\C$, the trivial representation.

A c-graph is a ribbon graph equipped with an assignment
of two words $w_{k} \in N(A)_{k}$,
$w_{l} \in N(A)_{l}$ to the bottom and top ends
of the open ribbons, together with an
assignment of an $A$\/-module $V$ to each ribbon.
($V$ is then called the colour
of the ribbon.)
These two assignments must be compatible in the sense
that the letters of $w_{k}$ and
$w_{l}$ corresponding to the ends of an open
ribbon must be equal to its colour,
and its direction has to be determined by the
signs $\eps_{i}$ according to the
following rule: if a ribbon end is labeled by a letter $V_{i}^{\eps_{i}}$,
then it is directed downwards (resp. upwards) if $\eps_{i} = +1$ (resp. -1).
Figure 3 shows an example of c-graph.

These definitions can be conveniently organised
into a category $\grc$ of c-graphs.
Its objects are the elements of $N(A) = \bigcup_{k} N(A)_{k}$,
and the morphisms
are the c-graphs. For example, the c-graph of figure 3 is a morphism
$V_{1}\Box (V_{2}\Box V_{3}^{*})\rightarrow (V_{1}\Box V_{3}^{*})\Box V_{2}$.
Notice that our convention is that a c-graph
is a morphism from the bottom to the
top. If a c-graph has no extremities of open ribbons at the bottom or the top,
then it is a morphism to or from $\C$. If it has no open ribbons at all,
we say that it is a closed c-graph.
We stress that the bottom and top objects, including the location of
parentheses, are essential parts of the definition of a morphism.
This is illustrated in figure 4.
\subsection{The functor $F$}
Our aim is to define a functor $F$ from $\grc$ to the category ${\rm Rep}(A)$
of
finite-dimensional representations of $A$, whose objects are finite-dimensional
$A$\/-modules, and morphisms are intertwiners. If $w \in N(A)$ then
$F(w)$ is the $A$\/-module obtained by replacing all formal products $\Box$ by
tensor products $\te$, and for a c-graph $C : w \rightarrow w^{\prime}$,
$F(C)$ is an intertwiner $F(w) \rightarrow F(w^{\prime})$.
The image $F(C)$ of a closed c-graph $C : \C \rightarrow \C$ is then
a pure number, which is the essential ingredient of the invariants of
links and 3-manifolds which we construct later.
The definition
of $F(C)$ is based on the observation that any c-graph $C$ can be
built from a few elementary ones by gluing and juxtaposition. These elementary
c-graphs $I, X^{\pm}, U, D, \Phi$ are shown on figure 5.

Let us define more precisely what we mean by gluing and juxtaposition.
Suppose that
$C : w \rightarrow w^{\prime}$ and
$C^{\prime} : w^{\prime} \rightarrow w^{\prime\prime}$ are two c-graphs.
Then by gluing we mean the composition of morphisms in $\grc$,
$C^{\prime} \circ C : w \rightarrow w^{\prime\prime}$,
which is obviously
defined as in figure 6. It is important that the top $w^{\prime}$ of $C$ is
exactly equal to the bottom of $C^{\prime}$, including the location of
parentheses.

Juxtaposition in $\grc$ is a binary operation $\Box$. For $w\in N(A)_{k}$,
$w\in N(A)_{l}$, it is simply $w_{k} \Box w_{l} \in N(A)_{k+l}$. For
c-graphs $C : w \rightarrow w^{\prime}$,
$C^{\prime} : x \rightarrow x^{\prime}$, we define
$C \Box C^{\prime} : w \Box x \rightarrow w^{\prime} \Box x^{\prime}$
by placing them side by side, as in figure 7.

Observe that in $\grc$ there is a class of c-graphs
$\Psi_{w}^{w^{\prime}}$, entirely made of vertical lines, and such that $w$ and
$w^{\prime}$ can differ only in the location of parentheses.
In figure 8 we have displayed the case
$w = (V_{1} \Box (V_{2} \Box V_{3}^{*})) \Box V_{4}$,
$w^{\prime} = (V_{1} \Box V_{2}) \Box (V_{3}^{*} \Box V_{4})$.

The functor $F$ is required to have the following properties:
it is a covariant functor,
\beq
F(C^{\prime} \circ C) = F(C^{\prime}) \circ F(C),
\eeq
juxtaposition corresponds to tensor products:
\beq
F(C \Box C^{\prime}) = F(C) \te F(C^{\prime}),
\eeq
and the $\Psi$ graphs enjoy a ``fusion'' property, which states that whenever
$w, w^{\prime} \in N(A)_{k}$ differ only in the location of parentheses, but
are such that they have a part
$(V_{i}^{\eps_{i}} \Box V_{i+1}^{\eps_{i+1}}) = w^{(i)}$
in common, then
\beq
F(\Psi_{w}^{w^{\prime}}) = F(\Psi_{w_{\te}}^{w^{\prime}_{\te}}),
\eeq
where $w_{\te} \in N(A)_{k-1}$ is obtained from $w$ by replacing $\Box$ by
$\te$
in $w^{(i)}$.
The functor $F$ is then defined by its values on the elementary graphs
of figure 5: $I, X^{\pm}, U, D, \Phi$, as follows:
\beq
F(I_{V}) = \id_{V}
\eeq
\beq
F(X^{+}_{V,W}) = \rc_{V,W}
\eeq
\beq
F(X^{-}_{V,W}) = \rc_{V,W}^{-1}
\eeq
\begin{eqnarray}
F(U^{R}_{V}) (f \te x) & = & f(\alpha \, x), \;\;\; f\in V^{*}, \, x\in V,
\\
F(U^{L}_{V}) (x \te f) & = & f(S(\alpha) uv^{-1} x)
\end{eqnarray}
\beq
F(D^{R}_{V}) (1) = \sum_{j} \beta \cdot e_{j} \te e^{j},
\eeq
where $\{e_{j}\}$ is a basis of $V$, and $\{e^{j}\}$ the dual basis of $V^{*}$,
\beq
F(D^{L}_{V}) (1) = \sum_{j} e^{j} \te u^{-1} v S(\beta) \cdot e_{j},
\eeq
\beq
F(\Phi^{V_{1},V_{2},V_{3}}) = \phi^{V_{1},V_{2},V_{3}}.
\eeq
Notice that the r.h.s. of these equations
are all intertwiners, as they should be.
One has to show that $F$ is well-defined.
This means two things: that $F$ preserves
all relations coming from isotopy of ribbons,
and that the value of $F$ on any c-graph
is independent from the choices made in evaluating it, i.e. cutting it into
smaller pieces until one reaches a decomposition into elementary graphs.
Let us elaborate on this latter point, which is more subtle than in the case
of Hopf algebras.

We show first that $F(\Psi_{w}^{w^{\prime}})$
is well-defined. In view of the
fusion property, it is clear that $F(\Psi_{w}^{w^{\prime}})$ is built from
$\phi, \phim$, and the identity operator.
There are several ways to evaluate
$F(\Psi_{w}^{w^{\prime}})$, however Mac Lane's
``coherence'' theorem \cite{maclane}
states that they all give the same result since $\phi$
satisfies the pentagon identity.
The properties of quasi-Hopf algebras involving the counit $\eps$ guarantee
the well-definedness of the c-graphs containing $U$ or $D$.

To prove that $F$ depends only on isotopy classes of c-graphs,
it is enough to prove that the relations listed on figure 9 are preserved
\cite{reshet,rt2}, for all possible colorings and directions of ribbons.
The proof that $F$ preserves relations (a), (b) and (c) is very simple:
(a) amounts to eq. (\ref{snake1}) and (\ref{snake2}), (b) is trivial and
(c) is eq. (\ref{qyb}). It can be shown that
\beq
F(L^{+}_{V}) = F(L^{\prime +}_{V}) = \pi(v^{-1}),
\label{loop1}
\eeq
\beq
F(L^{-}_{V}) = F(L^{\prime -}_{V}) = \pi(v),
\label{loop2}
\eeq
where the c-graphs $L^{\pm}_{V}, L^{\prime \pm}_{V}$ are given on figure 10.
This implies that (d) is also respected.
Note that these two equations reflect the fact that the objects
we are dealing with are ribbons, see figure 2 for a graphical
interpretation of (\ref{loop1}). It is
instructive to evaluate $F(L^{+}_{V})$ to illustrate how the definitions
are used in practice. Breaking $L^{+}_{V}$ into pieces one finds:
\begin{eqnarray}
F(L^{+}_{V}) & = & (F(U^{R}_{V}) \te \id_{V}) \, (\phi^{V^{*},V,V})^{-1}
(\id_{V^{*}} \te \rc_{VV}) \, \phi^{V^{*},V,V} (F(D^{L}_{V}) \te \id_{V})
\nonumber\\
 & = & \sum_{j,k,p} \pi( R_{k} \, a_{p} \, Y_{j} \,
             u^{-1} v \, S(P_{k} \, X_{j} \, \beta) \,
  \alpha \, Q_{k} \, b_{p} \, Z_{j}).
\end{eqnarray}
\subsection{$q$\/-traces and $q$\/-dimensions}
Suppose $ C \ : w \rightarrow w $ is a c-graph with the same words on
top and bottom, where $w\in N(A)_k $. We define the closure
$\widehat{C}$ of $C$ by figure 11.
By construction, $F(C)\in {\rm End} F(w)$ is an intertwiner. We put:
\beq
\qtr F(C) = \tr_{F(w)} (F(C)\b S(\a ) u\vm).
\eeq
The main properties of this definition are
\beq
\qtr (F(C \circ C')) = \qtr (F(C' \circ C)),
\label{trcommut}
\eeq
where $C'$ is also a c-graph $w\rightarrow w$, and
\beq
F(\widehat{C}) = \qtr F(C).
\label{qclosure}
\eeq
The proof of (\ref{qclosure}) uses axiom (R4) of ribbon quasi-Hopf
algebras, (\ref{eqt}) and (\ref{prof}).
Consider first the case $ C\ :\ V_1 \Box V_2 \rightarrow
                                          V_1 \Box V_2 \  $.
Let $\Lambda = (\pi_1 \te \pi_2 \te \pi^{*}_2 \te \pi^{*}_1)
(\del \te \id \te \id)(\phi) (\phim \te 1)$. Then
\begin{eqnarray}
F(\widehat{C}) & = & F(U^{L}_{V_1})
 (\id \te F(U^{L}_{V_2}) \te \id) \Lambda^{-1}
 (F(C)\te \id_{V_2^{*}}\te \id_{V_1^*})
  \Lambda (\id \te F(D^{R}_{V_2}) \te \id) F(D^{R}_{V_1}) \nonumber \\
     & = & \sum_{i,j} \tr_{V_1 \te V_2}[(S(\a D_i N_j )u\vm A_i \te
              S(\a C_i M_j )u\vm B_i ) F(C)(K_j \b \te L_j \b )] \nonumber \\
     & = & \tr_{V_1 \te V_2} [ F(C)
             (\delta (S \te S)(\gamma_{21})(u\vm \te u \vm ) ]\nonumber  \\
     & = & \tr_{V_1 \te V_2}[ F(C)\del (\b)f^{-1}
(S\te S)(f_{21} \del '(\a ))(u\vm \te u \vm ) ]        \nonumber \\
     & = & \tr_{V_1 \te V_2}[ F(C)\del (\b S(\a )) f^{-1}
           (S\te S)(f_{21})(u\vm \te u\vm ) ]         \nonumber \\
     & = & \qtr F(C).
\end{eqnarray}
The general case follows by induction.

\noindent
Finally, we define $q$\/-dimensions by:
\beq
\dimq (V)=\qtr (\id_{V})= \tr_{V} (\pi (\b S(\a ) u \vm )).
\eeq
Applying (\ref{qclosure}) to the identity graph
shows that $q$\/-dimensions are multiplicative,
\beq
\dimq (V_1 \te V_2 ) = \dimq (V_1) \cdot \dimq(V_2).
\eeq
{\bf Remark.} It is possible to give an alternative formulation of
(R4), which perhaps will be more appealing to the reader, as it
takes exactly the same form as the corresponding axiom for ribbon
Hopf algebras. It is based on a computation of $\del(u)$: from
(\ref{lemmaSalphau}),
(\ref{eqt}) and (\ref{deltaprime})
one derives, provided $\alpha$ is invertible:
\beq
\del(u) = f^{-1} (S\te S)(\gamma_{21}^{-1} f_{21} )
           \sum_p (S\te S)(\del^{\prime} (b_p) ) \gamma
           \del (a_p).
\eeq
Using the properties of the functor $F$ one can reexpress this as :
\beq
\del(u) = f^{-1} (S\te S)f_{21} (u\te u)(R_{21} R_{12})^{-1}
\eeq
But since one can also show that
\beq
(S\te S)R =f_{21} R f^{-1},
\eeq
which implies
\beq
(S\te S)(R_{12} R_{21}) = f R_{21} R_{12} f^{-1},
\eeq
the expression for $ \del(S(u)) = f^{-1} (S\te S) \del^{\prime}(u) f  $
becomes:
\beq
\del(S(u)) = (R_{21} R_{12})^{-1} (S(u)\te S(u))(S\te S)f_{21}^{-1} f.
\eeq
This leads to
\beq
\del (S(u)u) = (S(u)u\te S(u)u ) (R_{21} R_{12})^{-2},
\eeq
in agreement with (R1) and
\beq
\del (v) = (v\te v) (R_{21} R_{12})^{-1}.
\label{deltav}
\eeq
This condition is the axiom of ribbon Hopf algebras, which has the same
graphical interpretation in the quasi-Hopf case. In other
words (\ref{deltav}) is equivalent to (R.4), provided $\alpha$ is invertible.
\subsection{Representations of the braid group}
Any representation $(\pi , V)$ of a quasitriangular quasi-Hopf
algebra leads to a representation of the braid group $B_n$ of $n$
strands. The images of the generators $b_i$, $i=1,...,n-1$ are the
following endomorphisms of
$(((V\te V)\te V)\te \cdots )\te V = V_{L}^{\te n} $
(all left parentheses at the beginning):
\bea
 b_1    &=&\check{R}_{12} \\
 b_i    &=&\psi_i^{-1}  \check{R}_{i,i+1} \, \psi_i, \;\;\; i>1
\eea
where
\beq
 \psi_i = \pi^{\te n} ( \del_{L}^{i-2} (\phi)\te 1^{\te n-i-1}  ).
\eeq
Here $ \check{R}_{i,i+1} $ acts on the
$i$\/-th and $i+1$\/-th spaces parenthesed together, $\Delta_{L} $
is defined for any $ n \ge 1 $ by
\beq
\Delta_{L} (a_1 \te \cdots \te a_n )= \Delta (a_1)\te a_2 \te
                                           \cdots \te a_n,
\eeq
and the notation $\del^{k}_{L}$ stands for
$\del_{L} \circ \del_{L} \cdots \del_{L}$ ($k$ times) for $k \ge 1$,
$\del_{L}^{0} = \id$.
For instance, in the case of $B_5$, $\check{R}_{34}$  is a morphism of
$ ((V\te V)\te (V \te V))\te V$ and
\beq
b_3 = \pi^{\te 5} ((\Delta\te\id \te \id )(\phim) \te 1)\,
\check{R}_{34} \,
         \pi^{\te 5} ((\Delta\te\id \te \id )(\phi) \te  1).
\eeq
The braid group defining relations:
\begin{eqnarray}
     b_i b_j &= &b_j b_i \ \ \hbox{for}\ \  |i-j|\geq 2
       \label{braid1}\\
b_i b_{i+1} b_i &= &b_{i+1} b_i b_{i+1}  \label{braid2}
\end{eqnarray}
both come from conservation of isotopy by the functor $F$,
(\ref{braid2}) being a graphical representation of the quasi-Yang-Baxter
equation (\ref{qyb}).
We would like to stress that this result is less obvious than a naive
look would suggest, because of the insertions of $\Delta_{L}^k(\phi)$
operators which ensure the possibility of gluing together the
generators contained in a word of the braid group. In other words the
properties of $F$ imply identities such as:
\bea  \del_{L}^{i-1}(\phi) (\del^{i-2}_L(\phim) \te 1)
&=&(\id^{\te i} \te \del )\del^{i-2}_L (\phim) (1^{\te i-1} \te \phi )
    \del^{i-2}_L (\id \te \del \te \id )(\phi)\nonumber\\
    \del^{i-2}_L (\phi) \del^{j-2}_L (\phim)
&=&(\id^{\te i-1} \te \del \te \id^{\te n-i-1} )\del^{j-3}_L (\phim)
                          \del^{i-2}_L (\phi) \nonumber
\eea
which are consequences of the pentagonal identity, and can
be proven directly, although they result from Mac Lane's
coherence theorem.

This representation of the braid group depends on the choice of
parentheses made in $V_L^{\te n}\  $. However other choices for tensoring
$V$  with itself n times lead to equivalent representations.
The above choice allows an easy embedding of $B_n$
into $B_{n+1}$ when adding a strand to the right.

Let us now restrict our attention to the case where $(\pi,V)$ is an irreducible
representation with $\dimq V\neq 0$. Set
\beq
\T_n (g)=(\dimq V)^{-n}
    \, {\qtr}_{V_{L}^{\te n} } (g)  \label{defTn}
\eeq
where $g\in B_n$. Due to (\ref{trcommut}), (\ref{loop1}) and (\ref{loop2}),
$\T_n$ is a Markov trace:
\bea
               \T_n (g_1 g_2 )&=&\T_n (g_2 g_1 )  \\
      \T_{n+1}(g b_n^{\pm 1} )&=&\tau_V^{\pm } \ \T_n (g),
\eea
where
$\tau_V^{\pm } = \pi(v^{\mp 1}) \, / \, \dimq V$. This trace extends to
$B_{\infty}$, for
\beq
\T_n (g) = \T_m (g)\ \ \  \hbox{if}\ \ m>n, \;\;g\in B_n \subset B_m.
\eeq
 From $\T_n$ one can build ambient isotopy invariants of links
\cite{kauf,salzub}.
\section{The algebra $\dog$}
In this section, we recall the definition of the quasitriangular quasi-Hopf
algebra $\dog$ \cite{dpr,roche}. Then we show that $\dog$
is a ribbon quasi-Hopf algebra,
and finally we study the invariants of links
all of whose components are coloured
by the regular representation of $\dog$,
showing that they are in fact invariants
of the 3-manifolds obtained by surgery on $S^{3}$ along those links.

The algebra $\dog$ is a quasi-Hopf deformation of $D(G)$, the double of the
algebra ${\cal F}(G)$ of functions on a finite group $G$. Its definition
involves a 3-cocycle $\om : G \times G \times G \rightarrow U(1)$, which
is a normalized cochain, i.e.
$\om(x,y,z)=1$ whenever one (or more) of the
three arguments $x,y,z$ is (are) equal to the unit element of $G$.
Recall that by definition, a 3-cocycle $\om$ satisfies:
\beq
\om(g,x,y) \,\om(x,y,z) \,\om(gx,y,z)^{-1}\,\om(g,xy,z)\,\om(g,x,yz)^{-1} = 1,
\label{co}
\eeq
for any $g,x,y,z \in G$.
As a vector space, $\dog = {\cal F}(G) \te \C[G]$,
where $\C[G]$ is the group algebra.
Its structure will be given in terms of its
basis $\elt{g}{h} = \delta_{g}\te h$, $g,h\in G$.
Here $\delta_{g}(x)=\delta_{g,x}$.
To avoid confusion we denote by $e$ the
unit element in $G$, and by $1 = \sum_{g\in G} \delta_{g}$ the unit of
${\cal F}(G)$.
Sometimes we will use the notation $\elt{1}{g} = 1\te g$.
The algebra and coalgebra structures in $\dog$ are
as follows:
\beq
\elt{g}{x} \cdot \elt{h}{y} = \delta_{g,xhx^{-1}} \;
\th{g}(x,y) \;\;\elt{g}{xy}
\eeq
\beq
\Delta(\elt{g}{h}) =
\sum_{xy=g}\; \ga{h}(x,y) \;\;\elt{x}{h} \te \elt{y}{h}
\eeq
where $\th{g}(x,y)$ and $\ga{h}(x,y)$ are given by:
\beq
\th{g}(x,y) = \om(g,x,y) \,\om(x,y,(xy)^{-1} g xy) \,\om(x,x^{-1}gx,y)^{-1}
\label{dtheta}
\eeq
\beq
\ga{x}(g,h) = \om(g,h,x) \,\om(x,x^{-1}gx,x^{-1}hx) \,\om(g,x,x^{-1}hx)^{-1}
\label{dgamma}
\eeq
and therefore $\th{g}(x,y)$ and $\ga{g}(x,y)$ are also equal to one, as soon
as one of $g,x,y$ is equal to $e$. The unit element is $\ele{1}\;$.
The elements $\phi$ and $R$ are as follows:
\beq
\phi = \sum_{g,h,k \in G} \om(g,h,k)^{-1} \;\;\ele{g} \te \ele{h} \te \ele{k}
\eeq
\beq
R = \sum_{g \in G} \ele{g} \te \elt{1}{g}.
\eeq
The pentagon identity for $\phi$ is equivalent to the 3-cocycle relation
(\ref{co}), and
the relations (\ref{dtheta}), (\ref{dgamma}) are equivalent to the
quasitriangularity of $R$, eq. (\ref{quasitriangle1}) and
(\ref{quasitriangle2}).
Using the 3-cocycle relation (\ref{co}), one can check the identities:
\beq
\th{g}(x,y) \,\th{g}(xy,z) = \th{g}(x,yz) \,\th{x^{-1}gx}(y,z)
\label{id1}
\eeq
\beq
\ga{x}(g,h) \,\ga{x}(gh,k) \,\om(x^{-1}gx,x^{-1}hx,x^{-1}kx)
  = \ga{x}(h,k) \,\ga{x}(g,hk) \,\om(g,h,k)
\label{id2}
\eeq
\beq
\th{g}(x,y) \,\th{h}(x,y) \,\ga{x}(g,h) \,\ga{y}(x^{-1}gx,x^{-1}hx)
  = \th{gh}(x,y) \,\ga{xy}(g,h).
\label{id3}
\eeq
These relations imply respectively that multiplication is associative,
comultiplication is quasi-coassociative, and that the coproduct is
a morphism of algebras.
The counit and the antipode are defined by:
\beq
\eps(\elt{g}{h}) = \delta_{g,e}
\eeq
\beq
S(\elt{g}{x}) = \th{g^{-1}}(x,x^{-1})^{-1} \ga{x}(g,g^{-1})^{-1}
\;\;\elt{x^{-1}g^{-1}x}{x^{-1}}
\eeq
and $\alpha,\beta$ by:
\beq
\alpha = 1, \;\;\; \beta = \sum_{g \in G} \la{g} \;\ele{g}\: ,
\eeq
where we have set
\beq
\la{g} = \om(g,g^{-1},g).
\eeq
Note that $\beta$ is invertible,
$\beta^{-1} = \sum_{g \in G} \la{g}^{-1} \;\ele{g} = S(\beta)$,
and also that (\ref{co})
implies:
\beq
\la{g^{-1}} = \la{g}^{-1}.
\eeq
 From (\ref{dtheta}) and (\ref{dgamma}) one finds:
\beq
\th{g}(g,g^{-1}) = \ga{g}(g^{-1},g) =
\th{g}(g^{-1},g) = \ga{g}(g,g^{-1}) = \la{g}.
\eeq
Now we claim that for any $a\in\dog$,
\beq
S^{2}(a) = \beta^{-1} a \beta.
\label{s2beta}
\eeq
To prove this, one computes explicitly the action of $S^{2}$ on the
basis $\elt{g}{x}$ using (\ref{id1}), (\ref{id2}) and (\ref{id3}). An
immediate corollary of (\ref{s2beta}) is that $v \in \dog$ defined by
\beq
v = \beta u,
\label{defv}
\eeq
is central. We now show that $v$ defines a ribbon structure on $\dog$.
Remark that (\ref{defv}) implies that $\qtr (.) = \tr (.)$ and
$\dimq (.) = \dim (.) \neq 0$.
The proof of (R1), (R2) and (R3)
consists only of direct computations, and we omit the details. The reader
will check that:
\beq
u = \sum_{g \in G} \la{g}^{-2} \;\elt{g}{g^{-1}}
\eeq
\beq
S(u) = \sum_{g \in G} \elt{g}{g^{-1}}
\eeq
\beq
v = \sum_{g \in G} \la{g}^{-1} \;\elt{g}{g^{-1}},
\eeq
from which the equalities $v^{2} = u S(u)$, $S(v) = v$ and $\eps(v)=1$ follow.
It is also easy to compute explicitly:
\beq
f = \gamma = \sum_{g,h} \om(g^{-1},g,h) \,\om(h^{-1},g^{-1},gh)^{-1}
\;\; \ele{g} \te \ele{h}
\eeq
\beq
\delta = \sum_{g,h} \la{g} \,\la{h} \,\om(g,h,h^{-1}g^{-1})
\,\om(h,h^{-1},g^{-1})^{-1} \;\; \ele{g} \te \ele{h}
\eeq
thus (R4) is equivalent to the following identity:
\beq
\la{x} \,\la{y} \,\la{xy}^{-1} = \om(xy,y^{-1},x^{-1})
\,\om(y^{-1},x^{-1},x) \,\om(y^{-1}x^{-1},x,y)^{-1} \,\om(x,y,y^{-1})^{-1},
\eeq
which is implied by the 3-cocycle relation (\ref{co}).

\noindent
{\bf Remarks.}
1. The algebra $\dog$ is semisimple, i.e. all representations
are completely reducible. The proof of this is parallel to the standard
proof that $\C[G]$ is semisimple \cite{freed}: let $p$ be a projector on an
invariant subspace, and consider
\beq
p_{0} = \ordm{1} \sum_{g,x \in G} \ga{g}(x,x^{-1}) \;
S \left( \elt{x}{g} \right) \; p \; \elt{x^{-1}}{g} \: .
\eeq
Here $\ord$ is the order of $G$.
Then $p_{0}$ is a projector and an intertwiner. Hence the complementary
subspace ${\rm Ker} \, p_{0}$ is invariant.

\noindent
2. The ribbon invariants of closed c-graphs depend only on the cohomology
class of $\om$ in $H^{3}(G,U(1))$. Recall that $\om^{\prime}$ is equivalent
to $\om$ if they differ by a coboundary $\delta\eta$, where
$\eta : G \times G \rightarrow U(1)$ is a normalized cochain, and
\beq
\delta\eta(x,y,z) = \eta(y,z) \,\eta(xy,z)^{-1} \,\eta(x,yz) \,\eta(x,y)^{-1}.
\eeq
Now the element $f_{\eta}$ defines a twist of $\dog$, where
\beq
f_{\eta} = \sum_{g,h \in G} \eta(g,h) \;\; \ele{g} \te \ele{h} \: .
\eeq
The twisted algebra is isomorphic to $D^{\om\delta\eta}(G)$.
Since twists preserve
equivalence classes of representations, our claim on closed c-graphs
follows, because their invariants are traces on representations.

In the sequel we shall consider the invariants of c-graphs all of whose
ribbons are coloured by the (left) regular representation of $\dog$.
Let us call those graphs {\em regular} c-graphs.
Recall that the regular representation is
the representation on the space $\dog$, where the
algebra acts by left multiplication.
We will show that invariants of closed regular c-graphs are in fact
invariants of the 3-manifolds which they define by surgery, and
conjecture that these
3-manifolds invariants are equal, up to a normalisation factor, to the
partition functions of Dijkgraaf and Witten \cite{dw}. We will give a number
of arguments supporting this conjecture.

As a preliminary step, we give the values of $F$ on the elementary regular
c-graphs. We find
\beq
\rc \left( \elt{g_1}{x_1} \te \elt{g_2}{x_2} \right) =
\th{g_1 g_2 g_1^{-1}}(g_1,x_2) \;\;
\elt{g_1 g_2 g_1^{-1}}{g_1 x_2} \te \elt{g_1}{x_1}
\label{xplus}
\eeq
\beq
\rc^{-1} \left( \elt{g_1}{x_1} \te \elt{g_2}{x_2} \right) =
\th{g_2^{-1} g_1 g_2}(g_2^{-1},x_1) \,\th{g_1}(g_2,g_2^{-1})^{-1} \;\;
\elt{g_2}{x_2} \te \elt{g_2^{-1} g_1 g_2}{g_2^{-1} x_1}
\label{xminus}
\eeq
\beq
F(\Phi) \left(\elt{g_1}{x_1} \te \elt{g_2}{x_2} \te \elt{g_3}{x_3}\right) =
\om(g_1,g_2,g_3)^{-1} \;\;
\elt{g_1}{x_1} \te \elt{g_2}{x_2} \te \elt{g_3}{x_3}
\eeq
\beq
v \cdot \elt{g}{x} = \om(g,g^{-1} x,x^{-1} g x)^{-1} \;\; \elt{g}{g^{-1} x}
\label{vp1}
\eeq
\beq
v^{-1} \cdot \elt{g}{x} = \om(g,x,x^{-1} g x) \;\; \elt{g}{gx}
\label{vm1}
\eeq
Let $\{\psi_{g,x}\}$ be the dual basis of $\{\elt{g}{x}\}$. Then
(see figure 5):
\beq
F(U^{L}_{reg}) \left(\elt{g_1}{x_1} \te \psi_{g_2,x_2}\right) =
\la{g_1}^{-1} \delta_{g_1,g_2} \, \delta_{x_1,x_2}
\label{ul}
\eeq
\beq
F(U^{R}_{reg}) \left(\psi_{g_1,x_1} \te \elt{g_2}{x_2}\right) =
\delta_{g_1,g_2} \, \delta_{x_1,x_2}
\label{ur}
\eeq
\beq
F(D^{L}_{reg}) (1) = \sum_{g,x} \; \psi_{g,x} \te \elt{g}{x}
\label{dl}
\eeq
\beq
F(D^{R}_{reg}) (1) = \sum_{g,x} \la{g} \; \elt{g}{x} \te \psi_{g,x} \: .
\label{dr}
\eeq
To define 3-manifolds invariants we need first to recall the definition of
surgery on a link in $S^{3}$ \cite{rolf}.
We consider {\em framed links} $(L,f)$, where
$L = L_{1} \cup L_{2} \cup\ldots\cup L_{n}$ is
an oriented link in $S^{3}$ and
$f=(f_{1},\ldots,f_{n})$ are integers.
One can think of $(L,f)$ as being a ribbon graph
with an annulus $C_{i}$ corresponding to each $L_{i}$
such that the linking number
$\lk(\partial C^{+}_{i},\partial C^{-}_{i})$ of its two boundary components
$\partial C^{\pm}_{i}$ is equal to $f_{i}$.
Or one can draw a planar projection of
$L_{i}$ and compute its {\em writhe} \cite{kauf}:
\beq
\sum_{\textstyle\mbox{self-crossings} \,\, c} w(c)
\eeq
where $w(c)$ is defined by the rule: $w(X^{\pm}) = \pm 1$, the
symbols $X^{\pm}$ being the two crossings of figure 5.
This quantity is independent of the direction of $L_{i}$.
By inserting the appropriate number of loops $L^{\pm}$ (figure 10) we then
adjust the writhe so that it coincides with $f_{i}$.

Now we obtain a manifold $M_{L,f}$ from surgery on
$S^{3}$ as follows: we remove from
$S^{3}$ a tubular neighbourhood $U_{i}$ of each $L_{i}$.
Let $\mu_{i}$ be a meridian
on $\partial U_{i}$, i.e. a loop which is contractible in $U_{i}$, with
$\lk(\mu_{i}, L_{i})= +1$, and let $\lambda_{i}$ be a longitude, i.e. a loop on
$\partial U_{i}$, which is homologically trivial in $S^{3} - U_{i}$ with
$\lk(\lambda_{i}, L_{i})=0$. Consider a diffeomorphism
$h$ of $\bigcup_{i}\partial U_{i}$
such that $\mu_{i}$ is mapped to $J_{i} = \lambda_{i} + f_{i}\mu_{i}$
for each $i$.
Glue $U_{i}$ with $S^{3}-U_{i}$ using $h$,
identifying $\mu_{i}$ on $\partial U_{i}$
with $J_{i}$ on $\partial(S^{3}-U_{i})$.

The data $(L,f)$ is called a surgery
presentation of the manifold $M$ when $M$ is diffeomorphic to $M_{L,f}$.
In fact, every compact 3-manifold $M$ is diffeomorphic to some $M_{L,f}$,
in general there are even many distinct surgery presentations of a given
manifold (see below). We claim that
\beq
\cf(M_{L,f}) = \ordm{n} F(C_{L,f})
\label{invar}
\eeq
where $n$ is the number of components of $L$,
and $C_{L,f}$ is the regular c-graph
determined by $(L,f)$, is an invariant of the 3-manifold $M_{L,f}$, i.e. it is
independent of the surgery presentation $(L,f)$.
To prove this one can appeal to
a theorem of Kirby, Fenn and Rourke \cite{moves}, which says that $M_{L,f}$ is
diffeomorphic to $M_{L',f'}$ if and only if $(L,f)$ and $(L',f')$ are
related by a finite sequence of ``Kirby moves'' (see also Rolfsen \cite{rolf}).
Kirby moves are shown on figure~12. The most general move is
(12c), where a part of a framed link, containing $p$ vertical lines
intersecting transversally a two-dimensional disc bounded by a circle with
framing $\pm 1$, is replaced by $p$ parallel lines forming a composite loop
as indicated, or equivalently one performs a full twist on the $p$ lines and
the framing of each line changes by $\mp 1$. The circle on the left disappears
completely, so the number of components of the original link decreases by one.
Two important special cases are $p=0$ and $p=1$. When $p=0$ the Kirby move
simply consists in removing from the link an unknotted circle
(12a) with framing $\pm 1$, which is not linked to the other components.
Figure (12b) displays the case $p=1$.

It is easy to verify that $\cf$ evaluated for the two circles of figure
(12a) is equal to 1, using (\ref{vp1}), (\ref{vm1}),
and the rules (\ref{ul}-\ref{dr}).
This means that $\cf(S^3) = 1$, as
surgery on a circle with framing $\pm 1$ gives back $S^3$.
Notice that this
defines also our normalization of $\cf$, which is different than the one of
\cite{dw}, where they choose instead to normalize the invariant by requiring it
to have the value 1 on $S^2 \times S^1$. Our choice, which is the same as in
\cite{rt1}, ensures the multiplicativity under connected sums:
$\cf(M_1 \# M_2) = \cf(M_1) \cf(M_2)$.

For the proof of invariance of $\cf$ under a
general Kirby move, we will need the value of $F(C)$
for the c-graph $C$ of figure 13,
where $(\pi,V)$ is an arbitrary finite-dimensional representation:
\begin{eqnarray}
F(C) \,y & = & \sum_{g,x,h,k} \om(g^{-1},\, g,\, g^{-1}hg)^{-1}
\,\om(g^{-1},\, h,\, g) \,\om(h,\, k,\, g)^{-1}
\,\th{k}(h^{-1},\, h)^{-1}\nonumber\\
& & \;\;\;\;\;\;\;\pi(\elt{h}{g}) y \te r^{*}(\elt{k}{h^{-1}}) \psi_{g,x}
\te \elt{g}{x}
\label{schmalz}
\end{eqnarray}
where $y \in V$ and $r^{*}$ is the dual of the regular representation.
The proof that (\ref{invar}) is invariant under any Kirby move
rests on the following arguments: first we have a very useful graphical
interpretation of quasitriangularity, equations (\ref{quasitriangle1}) and
(\ref{quasitriangle2}) given by figure 14. Of course, we may iterate this
identification many times, thereby allowing us to ``fuse'' an arbitrary
number of lines in a crossing, preserving the location of parentheses.
Thus the invariant of the regular c-graph on the l.h.s. of (12c) is
equal to the invariant of the c-graph on the left of (12b), but now the
line which pierces the disc is coloured by a $p$-fold tensor product
of the regular representation with itself, while the boundary of the disc
is coloured by the regular representation. Now for {\em any} finite-dimensional
representation $(\pi,V)$ colouring the vertical line on the left of (12b),
with the $\pm 1$\/-framed circle coloured by the regular representation,
equation (\ref{schmalz}) implies that
the value of the corresponding invariant is
\beq
\ord \, \pi(v^{\pm 1}) .
\eeq
Since $\alpha=1$, we can apply equation (\ref{deltav}) of the remark
at the end of section 4.4, whose graphical content is the equality of
$F(L^{\prime \,\mp}_{V})$, where $V$ is the $p$-fold tensor product
mentioned before, with the r.h.s. of (12c). This concludes the proof
of the invariance of (\ref{invar}).

Note that the regular representation and its dual are equivalent. The reader
can check that
\beq
\psi_{g,x} \mapsto \ga{x}(g^{-1},g) \;\elt{g^{-1}}{x}
\eeq
defines an intertwiner. Thus, $\cf$ is independent of the directions of the
components of the link in the surgery presentation.

Now we state our conjecture, which is that $\cf(M)$ is, up to
the difference in normalization which we mentioned before, equal to the
partition function $Z(M)$ of \cite{dw}. Let us briefly summarize the definition
of $Z(M)$. Here $M$ will be a compact, connected, closed, oriented 3-manifold.
One could also treat the case of manifolds with boundaries,
but we shall refrain
ourselves from doing that for the sake of simplicity. Let us consider principal
fiber bundles $E \rightarrow M$ with structure group $G$. Since $G$ is
finite, the total space $E$ is just a finite covering of $M$ upon which
$G$ acts freely, the number of sheets being $\ord$, and $M \approx E/G$. It is
clear that these coverings or $G$\/-bundles are labeled by the group
homomorphisms $\rho : \pi_{1} M \rightarrow G$. The set
${\rm Hom}(\pi_1 M,G)$ of these homomorphisms is finite, and plays the role
of the set of gauge field configurations sectors in this topological
``Chern-Simons theory with finite gauge group''.
Let $E_G \rightarrow B_G$ be the universal
$G$\/-bundle. For any $G$\/-bundle $E \rightarrow M$ there is a bundle map
defined by the commutative diagram
\beq
\begin{array}{ccc}
E & \longrightarrow & E_G \\
\downarrow & & \downarrow  \\
M & \longrightarrow & B_G
\end{array}
\label{classe}
\eeq
which is unique up to homotopy.
The space $B_G$ is an Eilenberg-Mac Lane complex
$K(G,1)$, i.e. $\pi_1 B_G = G$, $\pi_n \, B_G = 0$, $n \ge 2$. The total
space $E_G$ is a contractible space with a free $G$ action. The map
$M \rightarrow G$ of
(\ref{classe}) is called a classifying map, because a $G$\/-bundle
$E \rightarrow M$ is uniquely characterized by this map. Therefore one may
identify $\rho\in {\rm Hom}(\pi_1 M,G)$ with a classifying map. In order to
define an action for the gauge field $\rho$ we interpret the 3-cocycle $\om$
as follows~: let $H^{*}(B_G,\Z)$ denote
the singular cohomology (this is {\em not}
the De Rham cohomology of differential forms,
see e.g. \cite{bott}). It is known that
$H^{*}(B_G,\Z) = H^{*}(G,\Z)$, where the r.h.s. is the group cohomology
(in fact this was the way Eilenberg and Mac Lane defined group cohomology
at the beginning). By the standard long exact sequence
argument one deduces that
$H^{k}(B_G,\R / \Z) = H^{k+1}(B_G,\Z)$. We choose a representative 3-cocycle
$\tilde{\om}$ in $H^{3}(B_G,\R / \Z)$. Then $\tilde{\om}$ is related to
$\om$ in (\ref{co}) by the exponential map
$\exp(2\pi i . ) : \R/\Z \rightarrow U(1)$. We regard $\R/\Z$ as an additive
group and $U(1)$ as a multiplicative one (of course they are isomorphic).
Usually one defines cohomology with an additive group of coefficients, as
opposed to the definition (\ref{co}) of $\om$, which uses multiplicative
coefficients. The action of a gauge configuration $\rho : M \rightarrow B_G$ is
\beq
\ca(\rho) = \langle\rho^{*}(\tilde{\om}), [M]\rangle \,\in \R/\Z,
\eeq
where $[M]$ is the 3-cycle
in singular homology given by the sum of
all 3-simplices (tetrahedra) in $M$. Hence the partition function is defined
by the following functional ``integral'', which is a finite sum:
\beq
Z(M) = \ordm{1}
\sum_{\rho\in {\rm Hom}(\pi_1 M,G)} \exp(2\pi i \ca(\rho)).
\label{zm}
\eeq
One can give a combinatorial definition
of $Z(M)$, a ``state model'' formulation
in the terminology of Kauffmann, as follows~:
take a triangulation of the oriented manifold
$M$, and assign an element of $G$ to each edge, such that the product
$g_1 g_2 g_3$ of elements corresponding to a triangle with the
induced orientation is equal to the identity. Also identify
an edge with positive orientation equipped with $g\in G$ to the same edge with
negative orientation, equipped with $g^{-1}$. Such an assignment is called
a state $\rho$ of the model. The partition function $Z(M)$
will be a sum over the
states of the Boltzmann weights of these states.
The weight $W(\rho) = \exp(2\pi i \ca(\rho))$ of a state is
\beq
W(\rho) = \prod_{t\in T} W_t
\eeq
where $T$ is the set of
all tetrahedra in $M$, and
\beq
W_t = \om(g,h,k)
\label{tetra}
\eeq
for the tetrahedron depicted in figure 15. The orientation of $M$ is
given by fixing the order of enumeration of the vertices for any
tetrahedron to be $(a,b,c,d)$ as in this figure. The
triangulation of the manifold $-M$ with
the opposite orientation is obtained by applying an odd permutation
of the vertices. In this case the weight (\ref{tetra}) of every
tetrahedron has to be changed according to $W_t \mapsto W_t^{-1}$.

Thus we see that the value of $Z(M)$ can be computed from a triangulation of
$M$, whereas $\cf(M)$  is computed from a surgery presentation.
This is why it is
not straightforward to show that the two are equal (up to a constant factor).
The general form of $\cf(M)$ is
\beq
\cf(M_{L,f}) = \ordm{n} \sum_{g_1,\ldots,g_N,x_1,\ldots,x_N \in G}
(\prod\delta_{{\rm relations},e})  \;\Omega(g_1,\ldots,g_N,x_1,\ldots,x_N)
\label{fm}
\eeq
There is one pair $(g_i,x_i)$ for each minimum in the c-graph $C_{L,f}$
representing $(L,f)$. The relations appearing as $\delta$ functions are
the image under a homomorphism $\rho : \pi_1 M \rightarrow G$ of a
presentation of $\pi_1 M$. Only the $g_i$, not the $x_i$, occur in these
relations. This comes from the fact that the crossings
(\ref{xplus}), (\ref{xminus}) in regular c-graphs implement the relations
in the Wirtinger presentation of $\pi_1 (S^3 - L)$. The additional relations
in $\pi_1 M$ resulting from surgery come from the $x_i$~: in fact in the
computation of $F(C_{L,f})$, $\delta$ functions appear at each maximum of the
graph due to (\ref{ul}), (\ref{ur}). Relations involving both $g_i$ and
$x_i$ are thus produced, from which the $x_i$, which are only present in the
second $\delta$ function of (\ref{ul}) and (\ref{ur}), can be eliminated, at
the cost of producing the surgery relations of $\pi_1 M$. The first $\delta$
function of (\ref{ul}) and (\ref{ur}) contributes to the Wirtinger relations.
This was noticed independently in \cite{hennings}, where the case of a trivial
cocycle $\om$ is discussed. Notice that the phase $\Omega$ disappears if the
cocycle is trivial, so in this case the preceding argument is the proof that
$Z(M) = \cf(M) = | {\rm Hom}(\pi_1 M,G) |$, the number of $G$\/-bundles on
$M$. (cf. \cite{alcos} for examples)
But when the cocycle $\om$ is non-trivial, the phase $\Omega$ is there,
coming from the factors $\theta,\gamma,\om$ of the rules for evaluating
regular c-graphs. So the precise form of the conjecture is that
\beq
\Omega(g_1,\ldots,g_N,x_1,\ldots,x_N) = W(\rho)
\eeq
where $\rho\in {\rm Hom}(\pi_1 M,G)$ is defined by the preceding discussion.

In order to check that $\cf(M)$ has the correct properties predicted by our
conjecture, we have computed its values for the lens spaces
$L_{p,1}$ and $L_{pq-1,q} = L_{pq-1,p}$, $p,q \ge 1$
(see e.g. \cite{rolf} for the definition and
classification of lens spaces). The former is presented by surgery on one
unknotted circle with framing $p$, the latter by surgery on the (framed)
Hopf link (two unknotted circles with linking coefficient $+1$) with
framings $p$ and $q$. Here are the results:
\beq
\cf(L_{p,1}) = \ordm{1} \sum_{g,h} \delta_{g^p,e}
\prod_{j=0}^{p-1} \om(g,\,g^j h,\, h^{-1}gh)
\label{lp1}
\eeq
\begin{eqnarray}
\cf(L_{pq-1,q}) & = & \ordm{2} \sum_{g,h,k} \delta_{g^{pq-1},e} \,
\th{g}(g^{-p},h) \,\th{g^{-p}}(g,k) \nonumber\\
& & \prod_{m=1}^{p} \om(g, \, g^{-m}h, \, h^{-1}gh)
    \prod_{n=0}^{q-1} \om(g^{-p}, \, g^{1-np}k, \, k^{-1} g^{-p} k)
\label{lpq}
\end{eqnarray}
In general, $\cf(M)$ is a complex number.
It follows from the definition of $Z(M)$ that $Z(-M) = Z(M)^{*}$
(complex conjugate). Hence $Z(M)$ is real if there exists
an orientation-reversing diffeomorphism on $M$. By the conjecture,
$\cf(M)$ should have these same properties, and so we checked them
for the lens spaces whose invariants are given above; it is easy to
show from (\ref{vp1}) that
\beq
\cf(-L_{p,1}) = \ordm{1} \sum_{g,h} \delta_{g^p,e}
\prod_{j=0}^{p-1} \om(g,\,g^j h,\, h^{-1}gh)^{-1}
= \cf(L_{p,1})^{*} .
\eeq
It is known that $L_{2,1} = \R{\rm P}^3 = -\R{\rm P}^3$. Therefore
(\ref{lp1}) with $p=2$ should be a real number. A little exercise
with the 3-cocycle identity shows that indeed it is real, for any
$G$ and $\om$. Another instructive exercise is to check that the
expressions (\ref{lp1}) and (\ref{lpq}) are invariant under the
substitutions $\om \mapsto \om\delta\eta$, see remark 2 above. Note
that the action $\ca(\rho)$ of $Z(M)$ depends only on the cohomology
class of $\tilde{\om}$, since $\partial M = \emptyset$.

We have also made a direct verification of the conjecture in the
case of $L_{p,1}$, by computing $Z(L_{p,1})$ from a triangulation
using the state model definition given before. A triangulation
of $L_{p,1}$ can be obtained as follows: take $p$ tetrahedra with
vertices labeled $(a_i,b_i,c_i,d_i)$ and edges $(g_i,h_i,k_i)$ as
in figure 15, with $i\in \{1,2,\ldots,p\}$. First glue together
the faces $(a_i,b_i,d_i)$ and $(a_{i+1},b_{i+1},c_{i+1})$, then glue
together $(c_i,d_i,a_i)$ and $(c_{i+1},d_{i+1},b_{i+1})$,
for $1 \le i \le p$, where $p+1$ is identified with 1.
This gluing process imposes relations among the group elements
$(g_i,h_i,k_i)$ of the edges, which lead to the expression
(\ref{lp1}).

For the group $G = \Z_2$, there exists a unique
non-trivial $\om$ given by $\om(g,g,g) = -1$, where $g \neq e$.
In this case, one can evaluate explicitly (\ref{lp1}) and find
\beq
\cf(L_{p,1}) = \left\{
\begin{array}{ll}
1 + (-1)^{p/2} & p \;\;{\rm even,}\\
1 & p \;\;{\rm odd,}
\end{array}   \right.
\eeq
in agreement with the corresponding value of $Z(L_{p,1})$ computed
in \cite{dw}.

For the cyclic group $G = \Z_n$ of order $n$, there is also an
explicit formula \cite{ms,roche}
for (a representative of) the generator $\om$ of
$H^3(\Z_n, U(1))$, which is a cyclic group of order $n$:
\beq
\om(x,y,z) = \exp (\frac{2\pi i}{n^2} \bar{z}(\bar{x}+\bar{y}
  - \overline{x+y}))
\eeq
where $\bar{x}$ is the representative of $x$ in the set
$\{0,1,\ldots,n-1\}$.

Put $(n,p) = {\rm gcd}(n,p)$.
It is possible to show that
$\cf(L_{p,1})$ for $G = \Z_n$ is a Gauss sum:
\beq
\cf(L_{p,1}) = \sum_{g=0}^{(n,p)-1} {\rm e}^{2i\pi p g^2/(n,p)^2}\ .
\eeq
The evaluation of these sums is a standard topic in the literature,
see e.g. \cite{dave}. It would be interesting to study
the arithmetic properties of the invariants in general, but for the moment
we shall only remark that
for any finite group $G$ of order $\ord = N$,
and any compact, closed manifold $M$, $\cf(M)\in \Q(q)$, where
$q$ is a primitive $N$\/-th root of unity, since any
$\om\in H^3(G,U(1))$ satisfies $\om^N = 1$ \cite{hilton}.

Using (5.53) one can compare the
invariants of $L_{pq-1,1}$ and $L_{pq-1,p}$ for cyclic groups.
(Remember that $\pi_1 L_{p,q} = \Z_p$ for any $q$.)
With the help of a computer program we evaluated
the expressions (5.49) and (5.50) in
a few cases. The results we found are:
\begin{center}
\bo{(5,1) \;\;\;\sqrt{5}}\\
\bo{(5,2) \;-\sqrt{5}}\\
\bo{(7,1) \;  i \sqrt{7}}\bo{(7,2) \;    i \sqrt{7}}\\
\bo{(11,1) \; i \sqrt{11}}\bo{(11,2) \; -i \sqrt{11}}\bo{(11,3) \; i \sqrt{11}}
\end{center}
In this table, each box contains $(p,q)$ followed by the value of
$\cf(L_{p,q})$. The group $G$ is always taken to be $\Z_p$, and the
cocycle $\om$ given by (5.53). Two different boxes correspond to two
manifolds which are not homeomorphic. Two boxes are
put on the same horizontal level if they correspond to two manifolds
having the same homotopy type.
(The classification of lens spaces by homeomorphism
and homotopy types is given e.g. in \cite{rolf}.)
The last row of the table shows that $\cf(M)$ is able to distinguish
manifolds with the same homotopy type, {\em in some cases}. It is
perhaps interesting to mention that the Jones-Witten invariant is
able to distinguish $L_{7,1}$ and $L_{7,2}$
\cite{freed2}, in contrast with the
results of the table. But at this time, it cannot be ruled out that
$\cf(M)$ becomes a finer invariant for other groups and cocycles.

{\bf Acknowledgements.} We would like to thank B. Barth\'{e}l\'{e}my,
P. Degiovanni, M. Domergue, E. Mourre, V. Pasquier,
P. Roche  and A. Roger for valuable discussions.
We also thank G. Shore for a careful
reading of the manuscript.
%
%
%

%
%
%
\newpage
\section*{Figure captions}
\begin{enumerate}
\item A ribbon graph
\item Representing a ribbon by a single line
\item A c-graph
\item Two different c-graphs
\item The elementary c-graphs
\item Gluing
\item Juxtaposition
\item A $\Psi$ graph
\item Isotopy relations
\item The four loops
\item The closure of a graph
\item (a) unknotted, unlinked circles with framing ${\pm 1}$
may be deleted. (b) example of Kirby move (c) general Kirby move
\item A c-graph
\item Quasitriangularity: these two graphs have the same invariants
\item An oriented tetrahedron with edges labeled by group
elements
\end{enumerate}
\end{document}